\documentclass[aps,prd,preprint,nofootinbib]{revtex4}

\usepackage{graphicx}
\usepackage{epsfig}

\begin{document}

\title{Hadronic Production of the Doubly Charmed Baryon
$\Xi_{cc}$ with Intrinsic Charm}

\author{Chao-Hsi Chang$^{1,2}$, Jian-Ping Ma$^{2}$,
Cong-Feng Qiao$^{3}$ and Xing-Gang Wu$^{2,4}$}
\address{$^1$CCAST (World Laboratory), P.O.Box 8730, Beijing
100080, P.R. China\\
$^2$Institute of Theoretical Physics, Chinese Academy of Sciences,
P.O.Box 2735, Beijing 100080, P.R. China\\
$^3$Department of Physics, Graduate University,
Chinese Academy of Sciences, Beijing 100049, P.R. China\\
$^4$Department of Physics, Chongqing University, Chongqing 400044,
P.R. China}


\begin{abstract}
The effects of the intrinsic charm on the hadronic production of
$\Xi_{cc}$ are studied. By taking reasonable intrinsic charm
component into account, the change of the theoretical prediction on
the production of $\Xi_{cc}$ for LHC and Tevatron is small, but in
contrast it may enhance significantly for SELEX. The reason is that
the collision energy at LHC and Tevatron is so large that the
gluon-gluon fusion sub-process, which is irrelevant to intrinsic
charm, becomes dominant. But the situation for SELEX is quite
different. Our numerical results for SELEX show that by considering
all the contributions from various sub-processes, the predicted
cross-section may be enhanced by a factor so big as $10^2$ due to a
modulating intrinsic charm being taken into account. Therefore, the
hadronic production of $\Xi_{cc}$ at SELEX may be sensitive enough
in observing the intrinsic charm inside
the incident hadrons. \\

\noindent {\bf PACS numbers:} 12.38.Bx, 12.39.Jh, 13.87.Ce, 14.20.Lq

\end{abstract}

\maketitle

\section{Introduction}

SELEX Collaboration \cite{exp,exp2} has reported the observation of
the doubly charmed baryon $\Xi^+_{cc}$, which contains two charm
quarks. According to their measurement, the decay width and
production rate, nevertheless, are much larger than expected
\cite{comm,the,baranov,kiselev1,kiselev2}. The theoretical
predictions based on gluon-gluon fusions in
\cite{comm,the,baranov,kiselev1,kiselev2} are much smaller than the
measured cross-section by order of about $10^3$. Several possible
mechanisms for the hadronic production of $\Xi^+_{cc}$ are proposed,
the discrepancy between experimental data and theoretical
predications decreases but it is still there \cite{cqww} and it is
at order of about $10^2$. Since the baryon $\Xi_{cc}$, either
$\Xi^+_{cc}$ or $\Xi^{++}_{cc}$, is doubly charmed, here we will
highlight the intrinsic charm in the colliding hadrons through
studying the hadronic production of $\Xi_{cc}$ at different
situations (e.g. at LHC, Tevatron and SELEX).

In the hadronic production, the $\Xi_{cc}$ is produced through
scattering or annihilating or fusion of two initial partons inside
the incident hadrons. Although the common belief is that the
gluon-gluon fusion mechanism is important and even dominant in high
energy hadron collision, the other mechanisms may be still
substantial sometimes. For the other mechanisms, besides the light
partons, it is also possible to find certain importance of heavy
quarks inside the initial hadrons, whose possibilities are described
by the corresponding heavy quark distribution function. The lightest
heavy-flavored charm-quark, being as a parton, can be generated in
different ways inside the incident hadrons. The heavy quark (charm
or bottom) component in parton distribution functions (PDFs) can be
perturbatively generated by gluon splitting, and hence is named as
the `extrinsic' component according to Ref.\cite{cqww}. It can also
be generated non-perturbatively and appears at or even below the
energy scale of heavy quark threshold. This component is normally
called as `intrinsic' one according to
Refs.\cite{pumplin,brodsky,brodsky1,brodsky3,bhps}. The extrinsic
charm contribution to the $\Xi_{cc}$ hadronic production has been
studied in Ref.\cite{cqww} within the general-mass
variable-flavor-number (GM-VFN) scheme \cite{gmvfn0,gmvfn1,gmvfn2}.
It is shown that the extrinsic charm contribution is substantially
large at the energy range of SELEX experiment \cite{cqww}, i.e., the
extrinsic charm can raise the total hadronic cross-section of
$\Xi_{cc}$ by almost an order of magnitude in comparison with that
only from the gluon-gluon fusion mechanism. It is also interesting
to investigate how much the intrinsic charm may contribute in the
hadronic production, specially to see where and to what `degree' the
intrinsic charm component in concerned hadron can be determined. In
fact, the existence of the intrinsic heavy flavor components in a
proton has been supported by some experiments and theoretical
predictions. For instance, the intrinsic heavy flavors are adopted
to study the diffractive Higgs production in Ref.\cite{brodsky2},
where the authors have pointed out that a clear experimental signal
for Higgs can be observed due to the existence of intrinsic heavy
flavor components inside the nucleon. Bearing the problem of the
intrinsic charm inside a nucleon in mind, we will examine the
intrinsic charm effects on the $\Xi_{cc}$ hadronic production at
different experiment situations, i.e. at LHC, Tevatron and SELEX,
and see which is the best place to perform the measurements on the
intrinsic charms\footnote{Since there are some puzzles in hadronic
production of $J/\psi$ still, thus we think that any determination
on the intrinsic charm components of the PDFs of the incident
hadrons cannot be decisive.}.

In fact, the intrinsic charm quark content of proton is still not
well determined. In the latest versions of PDF, like CTEQ
\cite{cteq6hq}, GRV \cite{grv} and MRST \cite{mrst}, the heavy quark
components arise only perturbatively through gluon splitting, which
means they have the extrinsic nature. Although these PDFs are
determined through the global fitting, it can not rule out the
intrinsic component with certain possibility because the data for
fitting does not include those experiment results which are
sensitive to the intrinsic heavy quark reactions \cite{pumplin}. At
present, the probability magnitude of the intrinsic charm within
nucleon has not been well fixed yet. Nevertheless, we know it can
not be be large in order to comply with the global fitting result.
Upper-bounds for the probability of the intrinsic charm $A_{in}$
inside nucleon, i.e., the first moment of the intrinsic charm quark
PDF at a lower energy scale (e.g. $\mu^2\sim m_c^2$) is available:
according to Refs.\cite{pumplin,brodsky,brodsky1,brodsky3}, $A_{in}$
can not be much larger than $1\%$; and another one,
$(0.86\pm0.60)\%$, is obtained by analysis of $F_2^c$ data in deep
inelastic muon scattering on iron \cite{ic1,ic2}. The intrinsic
charm PDF can only be studied with some nonperturbative methods with
models. Recently, the operator product expansion method shows that
the probability for Fock states in light hadron to have an extra
heavy quark pair of mass $M_Q$ decreases as $\Lambda^2_{QCD}/M_Q^2$
in non-Abelian gauge theory \cite{Franz}. In the present work for
definiteness, we will take the BHPS model \cite{bhps} as a typical
example for the intrinsic charm to study the $\Xi_{cc}$ hadronic
production.

The paper is organized as follows. In Sec.II, we outline the
techniques for calculating the $\Xi_{cc}$ hadronic production under
the GM-VFN scheme. Specially, we will show how we obtain the
requested PDF of charm quark at the higher energy scale through
CTEQ6HQ plus the intrinsic charm component obtained by solving DGLAP
equation with proper initial boundary condition at low energy scale.
In Sec.III, we present the numerical results for the $\Xi_{cc}$
hadronic production. The final section is reserved for discussions
and summary.

\section{Calculation Techniques}

With QCD factorization and the GM-VFN scheme, the differential
cross-section for the inclusive hadronic production of  $H_1 + H_2
\to \Xi_{cc} +X$ can be written as the convolution
\begin{eqnarray}
\sigma&=&  f^{g}_{H_{1}}(x_{1},\mu) f^{g}_{H_{2}}(x_{2},\mu)
\otimes\hat{\sigma}_{gg\rightarrow
\Xi_{cc}}(x_{1},x_{2},\mu)\nonumber\\
&+& \Big\{\sum_{i,j=1,2;i\neq
j}f^{g}_{H_{i}}(x_{1},\mu)\Big[f^{c}_{H_{j}}(x_{2},\mu)
-f^{c}_{H_{j}}(x_{2},\mu)_{SUB} \Big] \otimes
\hat{\sigma}_{gc\rightarrow
\Xi_{cc}}(x_{1},x_{2},\mu)\Big\}\nonumber\\
&+& \Big\{\sum_{i,j=1,2;i\neq
j}\Big[\left(f^{c}_{H_{i}}(x_{1},\mu)
-f^{c}_{H_{i}}(x_{1},\mu)_{SUB}\right)
\cdot\Big(f^{c}_{H_{j}}(x_{2},\mu)-f^{c}_{H_{j}}(x_{2},\mu)_{SUB}
\Big)\Big] \nonumber\\
&&\otimes \hat{\sigma}_{cc\rightarrow
\Xi_{cc}}(x_{1},x_{2},\mu)\Big\}+ \cdots, \label{pqcdf0}
\end{eqnarray}
where $\mu$ is the renormalization (factorization) scale, the
ellipsis stands for contributions from higher order in $\alpha_s$
and from light quarks initiate processes. The later are much smaller
than those given explicitly in Eq.(\ref{pqcdf0}). $f_H^a (a=g,c)$ is
the PDF of the corresponding parton $a$ and $f^{c}_{H}(x,\mu)_{SUB}$
is defined in the GM-VFN scheme as
\begin{eqnarray}\label{subtraction}
f^{c}_{H}(x,\mu)_{SUB}& \equiv & f^{g}_{H}(x,\mu)\otimes
f^{c}_g(x,\mu)=\int^1_{x}\frac{dy}{y}f^{c}_g(y,\mu)
f^{g}_{H}\left(\frac{x}{y},\mu\right)
\end{eqnarray}
with
\begin{eqnarray} f^c_g(x,\mu) &=& \frac{\alpha_s(\mu)}{2\pi}
\ln\frac{\mu^2}{m^2_c}P_{g\to q}(x)=\frac{\alpha_s(\mu)}
{2\pi}\ln\frac{\mu^2}{m^2_c} \cdot \frac{1}{2}(1-2x+2x^2).
\end{eqnarray}

To be realistic we will consider only the $\Xi_{cc}$ with nonzero
transverse momentum $p_t$. Therefore, the partonic cross-sections
in Eq.(\ref{pqcdf0}) represent various parton processes at leading
order in $\alpha_s$. The $\hat{\sigma}_{gg\rightarrow \Xi_{cc}}$
stands for $g+g \to \Xi_{cc} + \bar c +\bar c$,
$\hat{\sigma}_{gc\rightarrow \Xi_{cc}}$ stands for $ g+c \to
\Xi_{cc} +\bar c$ and $\hat{\sigma}_{cc\rightarrow \Xi_{cc}}$ for
$c + c\to \Xi_{cc} +g$. These partonic cross-sections can be
further factorized if we take charm quark as heavy quark. In this
case, charm quarks inside the rest $\Xi_{cc}$ move with a small
velocity $v$. One can systematically expand the partonic
cross-sections in Eq.(1) in $v$ by using nonrelativistic QCD
(NRQCD) \cite{nrqcd} to separate perturbative and nonperturbative
effects. In this framework the partonic cross-section can be
expressed as \cite{majp}:
\begin{equation}
\hat{\sigma}_{ab\rightarrow \Xi_{cc}} = H (ab\to
(cc)[^3S_1]_{\bf\bar 3} ) \cdot h_3 +H (ab\to (cc)[^1S_0]_{\bf 6} )
\cdot h_1 +\cdots, \label{nrqcd10}
\end{equation}
where the ellipsis stands for the terms in higher orders of $v$. $H
(ab\to (cc)[^3S_1]_{\bf\bar 3} )$ or $ H (ab\to (cc)[^1S_0]_{\bf 6}
)$ is the perturbative coefficient for producing a $cc$ pair in
configuration of $^3 S_1$ and color ${\bf\bar 3}$, or $^1 S_0$ and
color ${\bf 6}$ respectively \cite{cqww}. The parameters $h_3$ and
$h_1$ characterize the transitions of a $(cc)[^3S_1]_{\bf\bar 3} $
pair and a $(cc)[^1S_0]_{\bf 6}$ pair into the produced $\Xi_{cc}$,
respectively. They are nonperturbative in nature and are just the
relevant matrix elements in NRQCD framework (see
Eq.(\ref{nrqcd10})). It has been pointed out in \cite{majp} that
these two parameters are at the same order of $v$, so for
convenience and to decrease the `freedom' of the prediction,
hereafter we assume $h_1$ is equal to $h_3$.

Now we turn to the part on how to deal with the additional
contributions from the intrinsic charm components to the charm PDF.
It is generally expected that the intrinsic heavy quark component in
PDFs is proportional to $\Lambda^2_{QCD}/M_Q^2$ with $M_Q$ being the
mass of the heavy quark \cite{Franz}. Therefore, in general the
heavy quark components are small, so we will treat the intrinsic
charm component as a small perturbation to the current determined
PDFs. We will also take the leading order DGLAP equation for its and
the relevant gluon component evolution. Hence, at leading order the
introduction of the intrinsic charm component will not affect the
PDFs of light quarks, but will affect the gluon's PDF. Under the
above approximation the charm- and gluon- PDFs can be written as:
\begin{eqnarray}
f_H^c(x,\mu)&=& f_H^{c,\;ex}(x,\mu)+f_H^{c,\;in}(x,\mu),\\
f_H^g(x,\mu)&=& f_H^{g,\;ex}(x,\mu)+f_H^{g,\;in}(x,\mu).
\label{apsg}
\end{eqnarray}
In the above equations, $f_H^{c,\;ex}(x,\mu)$ is the extrinsic charm
component that is already determined by the global fitting of
several groups. $f_H^{c,\;in}(x,\mu)$ stands for the intrinsic charm
component. The intrinsic charm component will also induce a small
change to the currently determined gluon PDF, which is denoted as
$f_H^{g,\;in}(x,\mu)$. Hence, the gluon PDF is the sum of the change
$f_H^{g,\;in}(x,\mu)$ and the currently determined gluon PDF
$f_H^{g,\;ex}(x,\mu)$.

The intrinsic charm component can not be calculated with pQCD. It
can be introduced only with nonperturbative methods at some lower
energy scale. Several models have been constructed for
$f^{c,\;in}_P(x,\mu)_{\mu=2m_c}$ in a proton \cite{pumplin,vvkov}.
The function $f^{c,\;in}_P (x,\mu)_{\mu=2m_c}$ from various models
constructed in Ref.\cite{pumplin} is close to that of the BHPS model
\cite{bhps} in shape, so we will take BHPS model as a typical one in
our numerical studies. In this model the intrinsic charm component
at $\mu =2m_c$ is parameterized as:
\begin{equation}
f^{c,\;in}_P(x,2m_c) = f^{\bar c,\; in}_P(x,2m_c) = 6 \xi
\left[6x(1+x) \ln x + (1-x) (1 + 10x + x^2) \right] x^2 \; ,
\label{bhps}
\end{equation}
where $P$ stands for the proton, the parameter $\xi$ is determined
by the first momentum of the distribution, i.e., the probability to
find a charm quark in total:
\begin{displaymath}
A_{in}\equiv \int_0^1 f_P^{c,\;in}(x,2m_c)\; dx=\xi\times 1\% \;.
\end{displaymath}
When $\xi=1$, it means that the probability for finding
$c/\bar{c}$-component in proton at the fixed low-energy scale $2m_c$
is $1\%$ as suggested in \cite{bhps,brodsky3}. In the following, we
will take a broader range $\xi\in[0.1,1]$ to do our discussions
\footnote{A quite smaller value of $A_{in}\sim 10^{-5}$ has been
suggested in Ref.\cite{vvkov}, since it is model dependent, we will
not take such a small value to do our calculation.}. The charm
content in an anti-proton is the same as that in a proton. Since we
will only deal proton-proton or proton-anti-proton scattering we
will suppress the subscript $H$ for PDFs and introduce shorthand
notations for those in Eq.(4):
\begin{equation}
f_H^{c,\;in} = f^{in}_c,\ \ \ f_H^{g,\;in} = f_g^{in},\ \ \
f_H^{c,\;ex} = f^0_c,\ \ \ f_H^{g,\;ex} = f_g^0, \ \ \ \ f_H^c =
f_c, \ \ \ \  f_H^g = f_g,
\end{equation}
where $H$ stands for a proton or anti-proton.

With the intrinsic component fixed at the initial scale $2m_c$ we
can obtain the distributions at any higher scale $\mu$ by
employing the DGLAP equation. Taking the leading order equation,
we use the the approximate method of Ref.\cite{apsolution} to
solve the DGLAP equations for $f^{in}_c(x,\mu)$ and
$f^{in}_g(x,\mu)$. The solutions are:
\begin{eqnarray}
f^{in}_c(x,\mu)&=& \int_x^1
\frac{dy}{y}\left\{f^{in}_c(x/y,2m_c)
\frac{[-\ln(y)]^{a\kappa-1}}{\Gamma(a\kappa)}\right\}+\nonumber\\
&& \kappa\int_x^1\frac{dy}{y} \int_y^1 \frac{dz}{z} \left\{
f^{in}_c(y/z,2m_c) \frac{[-\ln(z)]^{a\kappa-1}}
{\Gamma(a\kappa)}P_{\Delta c}(x/y)\right\}+{\cal
O}(\kappa^2)
\nonumber\\
f^{in}_g(x,\mu)&=& \frac{2\kappa}{a_g-a_c}\int_x^1
\frac{dy}{y}\int_{a_c}^{a_g}da\int_y^1
\frac{dz}{z}\left\{f^{in}_c(z,2m_c)
\frac{[-\ln(z)]^{a\kappa-1}}{\Gamma(a\kappa)}P_{c\to
gc}(x/y)\right\}+{\cal O}(\kappa^2),\label{intr}
\end{eqnarray}
with
\begin{eqnarray}
a_g& =& 6,  \ \ \ \ a_{c}=\frac{8}{3},\ \ \ \ \  \kappa=\frac
{2}{\beta_0} \ln \left(\frac{\alpha_s(2m_c)}{\alpha_s(\mu)}\right),
\ \ \ \ \ \beta_0=11-2n_f/3,
\nonumber\\
P_{\Delta c}(x) &=& \frac{4}{3}\left[\frac{1+x^2}{1-x}+\frac{2}{\ln
x}+\left(\frac{3}{2}-2\gamma_E\right)\delta(1-x)\right],
\nonumber\\
P_{c\to gc} &=&\frac{4}{3}\left[\frac{1+(1-x)^2}{x}\right]\,.
\label{intrg1}
\end{eqnarray}
where $n_f$ is the number of the flavor, and is taken to be $4$. It
should be noted that $P_{\Delta c}$ is not exactly the splitting
function $P_{c\to gc}$ \cite{apsolution}. With the above equations
we can obtain PDF's at any scale. To make the employed approximation
valid, it has been suggested in Ref.\cite{apsolution} that $\kappa$
given in Eq.(9,10) should be smaller that $0.3$. We have checked
this in our case with $\mu=M_t\equiv \sqrt{m_{\Xi_{cc}}^2+p_{t}^2}$.
For all ranges of $\mu$ used here $\kappa$ is smaller than $0.2$.
Since we include the intrinsic charm in PDF, the PDF of light quarks
and gluons will also be changed in order to satisfy the momentum sum
rule. This change can be expected to be at order of $\kappa^2$ and
leads to a violation of the momentum sum rule at $1\%$ level with
$A_{in}=1\%$. We will simply neglect this change since the change is
small.

\begin{figure}
\centering
\includegraphics[width=0.45\textwidth]{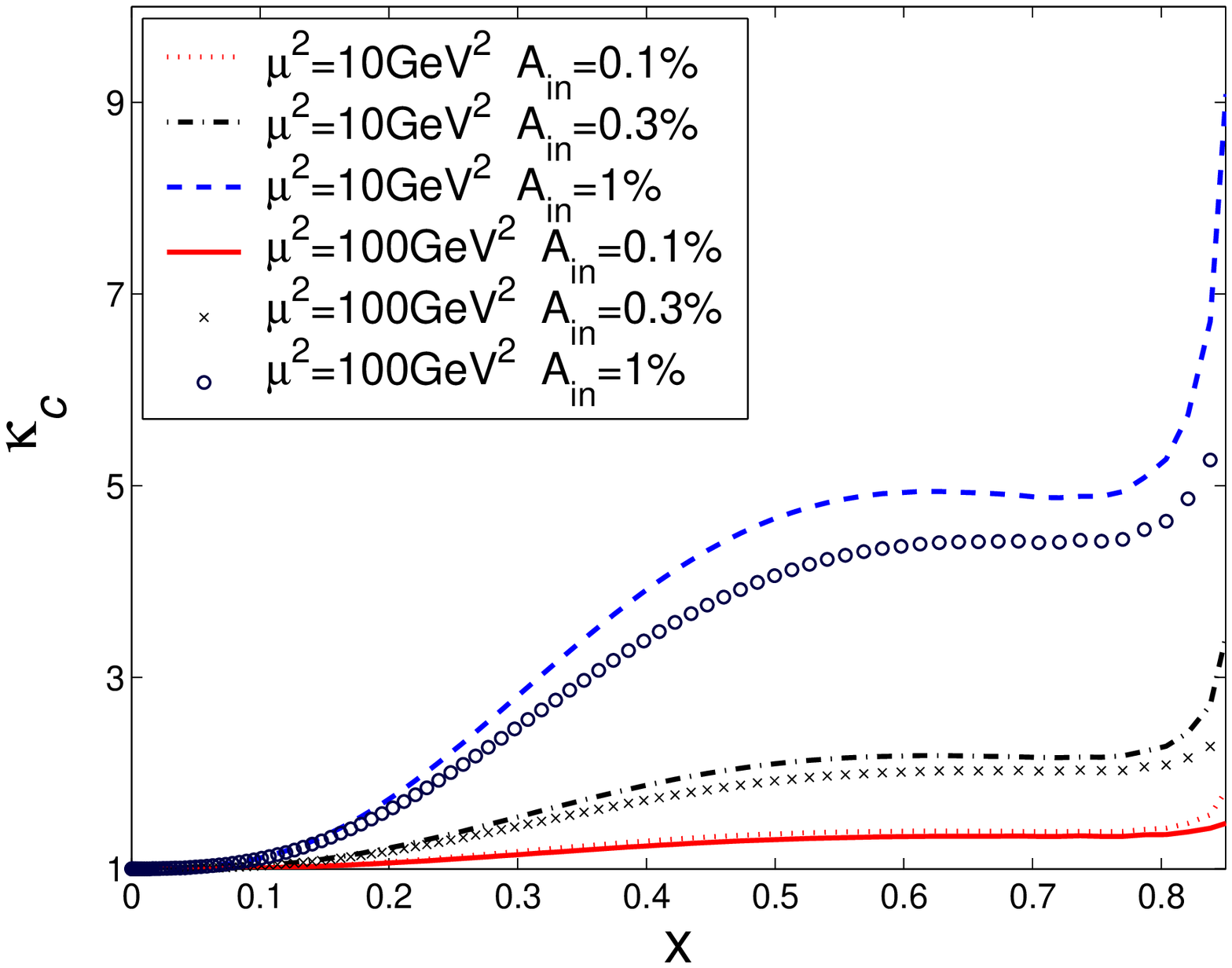}
\hspace{0.2cm}
\includegraphics[width=0.45\textwidth]{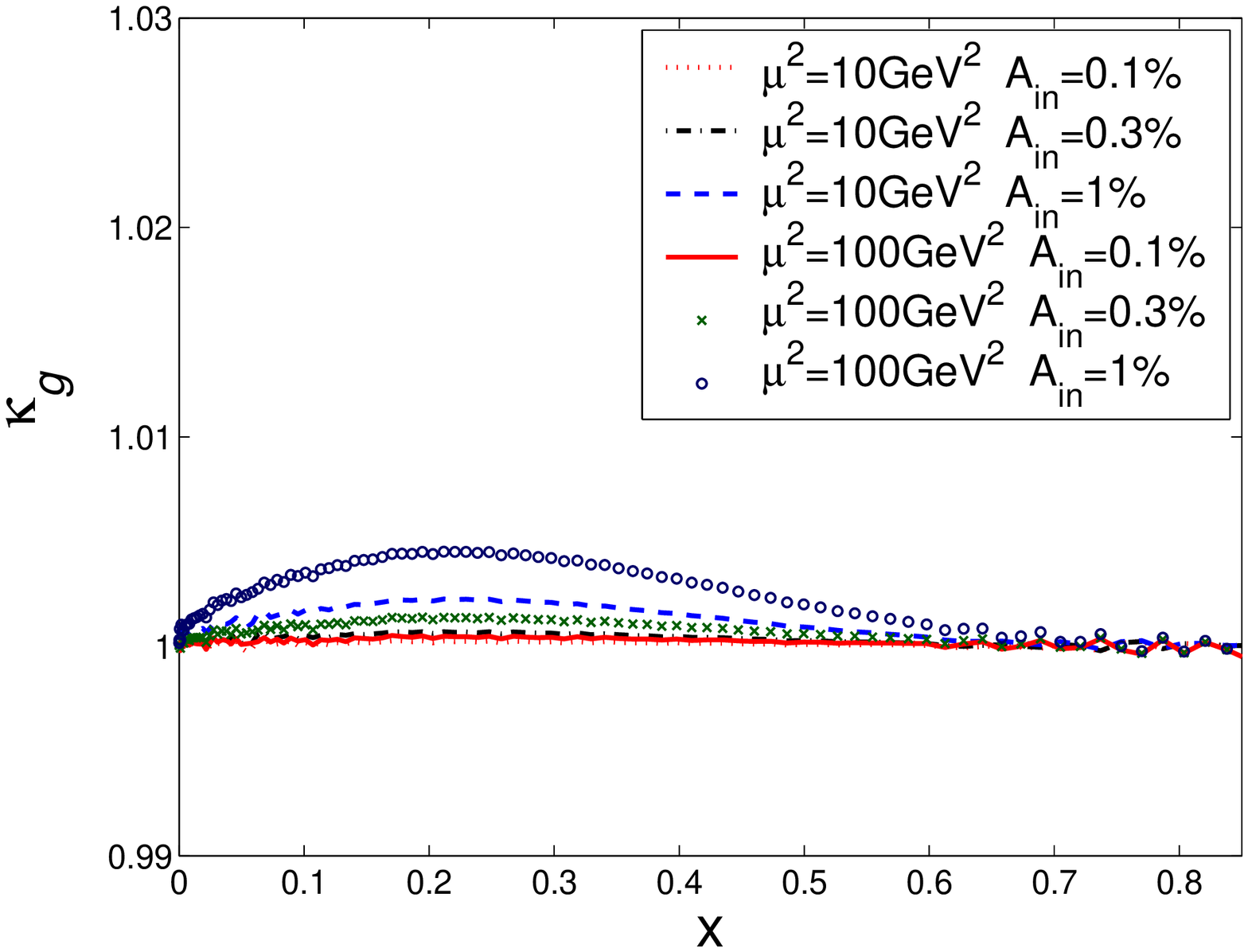}%
\caption{The ratio $\kappa_c=f_c(x,\mu)/f^{0}_c(x,\mu)$ (left) and
the ratio $\kappa_g=f_g(x,\mu)/f^{0}_g(x,\mu)$ (right). Two energy
scales and three typical normalization for the intrinsic charm,
$A_{in}=0.1\%$, $0.3\%$ and $1\%$, are taken respectively. }
\label{methodA}
\end{figure}

\begin{figure}
\centering
\includegraphics[width=0.49\textwidth]{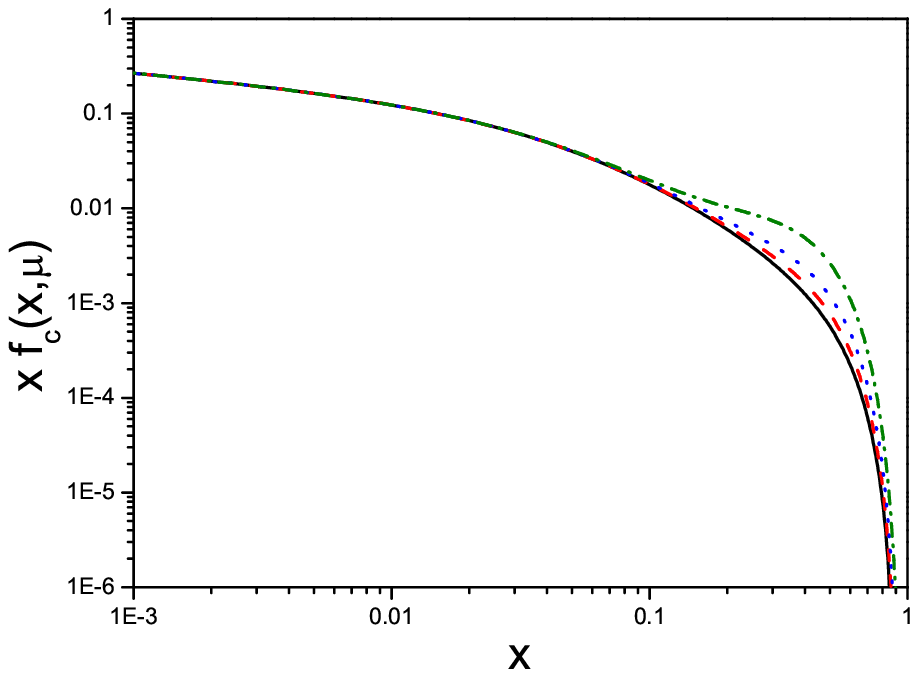}%
\hspace{0.2cm}
\includegraphics[width=0.49\textwidth]{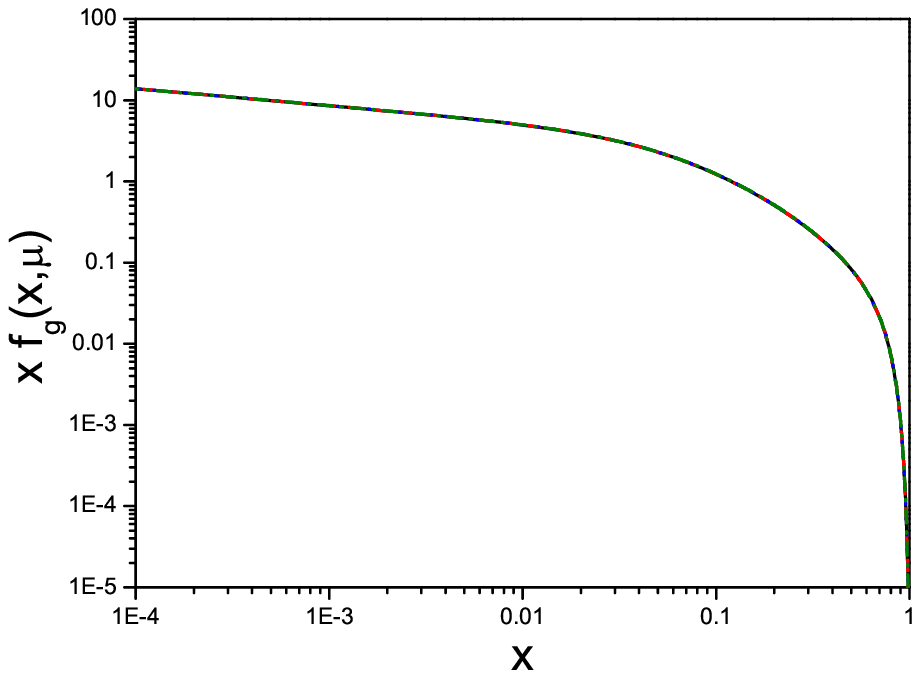}%
\caption{The charm PDF (left) and gluon PDF (right) at
$\mu^2=10\;{\rm GeV}^2$.  The solid line is for $xf^{0}_c(x,\mu)$ or
$xf^{0}_g(x,\mu)$. The dashed line, the dotted line and the
dash-dot line are for  $xf_c(x,\mu)$ or
$xf_g(x,\mu)$ with $A_{in}=0.1\%$, $0.3\%$ and $1\%$, respectively.}
\label{methodAn}
\end{figure}

To see how the intrinsic charm component modifies the PDFs, the
ratios $\kappa_c=f_c(x,\mu)/f^{0}_c(x,\mu)$ and
$\kappa_g=f_g(x,\mu)/f^{0}_g(x,\mu)$ are plotted in
Fig.\ref{methodA} at the energy scales of $\mu^2 = 10{\rm GeV}^2$
and $\mu^2 = 100{\rm GeV}^2$ with $A_{in}=0.1\%$, $0.3\%$ and $1\%$.
The distributions with the same parameters are plotted in
Fig.\ref{methodAn}. From Fig.\ref{methodA} one can see that the
intrinsic charm component changes the charm PDF significantly and
$\kappa_c$ increases as $x$ increases. When $x>0.8$ the change of
the charm PDF after including the intrinsic charm component is too
large to warrant the used approximations. Fortunately, the $x$ which
leads to the dominant contributions to differential cross section
falls in the intermediate region. Our numerical study also shows
that the PDFs in the region of $x\geq 0.8$ only give negligible
contributions to differential cross-sections. From Fig.1 and Fig.2
we can see that the change of the gluon PDF by including the
intrinsic charm component is indistinctive as expected.

\section{Numerical Results of Differential Cross-Sections}

In order to make numerical estimations on the differential
cross-sections we need to know the parameters $h_1$ and $h_3$.
Unfortunately, they are unknown. For simplicity, we will take $h_1 =
h_3$ as claimed in above. If one uses a nonrelativistic wave
function $\Psi_{(cc)}$ for the $^3S_1$ $cc$ pair in the color
${\bf\bar 3}$ state, then we have $h_3=|\Psi_{(cc)}(0)|^2$
\cite{majp}. We take $h_3=|\Psi_{(cc)}(0)|^2=0.039\;{\rm GeV}^{3}$
as in \cite{baranov}. The uncertainties from the values of $h_1$
and $h_3$ can be easily obtained, since they are overall factors for
the processes. For masses we take $m_{\Xi_{cc}}=3.50$~GeV and
$m^{eff}_c=1.75$~GeV. The energy scale $\mu$ is fixed to be the
transverse mass of $\Xi_{cc}$, i.e. $\mu=M_t\equiv
\sqrt{m_{\Xi_{cc}}^2+p_{t}^2}$, where $p_t$ is the transverse
momentum of the baryon. In dealing with the running coupling
$\alpha_s$ we take $n_f=4$ and $\Lambda^{(n_f=4)}_{QCD}=0.215$~GeV.
For all the numerical evaluations in the following we make use of
the CTEQ6HQ version of PDFs. The analytical and numerical results
are obtained by using well-established codes of BCVEGPY
\cite{bcvegpy} and FDC \cite{fdc}. Furthermore, one can conveniently
use the newly developed generator GENXICC to study the hadornic
production of $\Xi_{cc}$ \cite{genxicc}.

\subsection{Hadronic production of $\Xi_{cc}$ at SELEX}

\begin{table}
\begin{center}
\caption{The contribution of $\sigma_{ab}$ from different
sub-processes initialized by the partons $ab$ to the total cross
section (in pb) for the $\Xi_{cc}$ hadronic production at SELEX
with the cut of $p_{t}>0.2$~GeV.} \vskip 0.6cm
\begin{tabular}{|c||c|c|c||c|c|c|}
\hline - & \multicolumn{3}{|c||}{~~CTEQ6HQ($A_{in}=0$)~~}&
\multicolumn{3}{|c|}{~~$A_{in}=1\%$~~}\\
\hline\hline - & ~~$\sigma_{gg}$~~ & ~~$\sigma_{cc}$~~&
~~$\sigma_{gc}$~~ & ~~$\sigma_{gg}$~~& ~~$\sigma_{cc}$~~ &
~~$\sigma_{gc}$~~\\
\hline ~~$(cc)_{\bf\bar 3}[^3S_1]$~~ & ~~$4.03$~~  & ~~
$1.02\times10^{-3}$~~& ~~$102.$~~  &
~~$4.06$~~ & ~~$1.25\times10^{-2}$~~  & ~~$372.$~~ \\
\hline ~~$(cc)_{\bf 6}[^1S_0]$~~ & ~~$0.754$~~ &
~~$4.15\times10^{-5}$~~~ & ~~~$11.3$~~~ &
~~~$0.758$~~~& ~~~$5.01\times10^{-4}$~~~ & ~~~$40.9$~~~\\
\hline
\end{tabular}
\label{csA}
\end{center}
\end{table}

\begin{table}
\caption{The contribution rates of the sub-process $gc \to \Xi_{cc}$
in the different $x$ region in the charm quark PDFs with
$A_{in}=1\%$ and  $p_{t}>0.2$~GeV.} \vskip 0.6cm
\begin{tabular}{|c||c|c|c|c|c|}
\hline   {~~$0.0\leq x_c\leq 0.2$~~}&{~~$0.2\leq x_c\leq 0.4$~~}&
{~~$0.4\leq x_c\leq 0.6$~~}& {~~$0.6\leq x_c\leq 0.8$~~}& {~~$0.8\leq x_c\leq 1.0$~~}\\
\hline\hline   ~~$25\%$~~
 &~~$50\%$~~ & ~~$22\%$ ~~& ~~$3\%$ ~~& ~~$\sim 0$ ~~\\
\hline
\end{tabular}
\label{tabalpha}
\end{table}

The cross-sections receive contributions from different
sub-processes. To see how these sub-processes contribute to
$\Xi_{cc}$ production cross-sections we give our numerical results
of different sub-processes in TABLE.\ref{csA}, where we indicate
explicitly the initial parton states and the final $cc$ pair states.
From TABLE.\ref{csA} it is obvious that the intrinsic charm has the
most significant impact on the contribution from the sub-process
$gc\to \Xi_{cc}$. Also to note that this sub-process gives the
dominant contribution to the total cross-section with $A_{in}=0$ or
$A_{in}=1\%$. By taking $A_{in} =1\%$ the contribution is roughly
four times larger than that without intrinsic charm. Taking the
$g+c$ sub-process as an explicit example, we present how the
different regions of $x$ in the charm quark PDFs contribute to the
sub-process in TABLE.\ref{tabalpha}. From TABLE.\ref{tabalpha}. one
can see that the region $x>0.6$ gives tiny contribution and the main
contribution comes from $0.2< x <0.4$. We also note that in
TABLE.\ref{csA}. the contribution from the quark pair $c\bar c$ in
the configuration $(c\bar c)_{\bar 6} [^1S_0]$ are generally smaller
by a factor of about $10^{-1}$ than that in the configuration
$(c\bar c)_{\bf \bar 3} [^1S_0]$. It should be noted that the
contribution from $(c\bar c)_{\bar 6} [^1S_0]$ is zero if we take
$h_1=0$ as indicated in Eq.(4).

\begin{figure}
\centering
\includegraphics[width=0.46\textwidth]{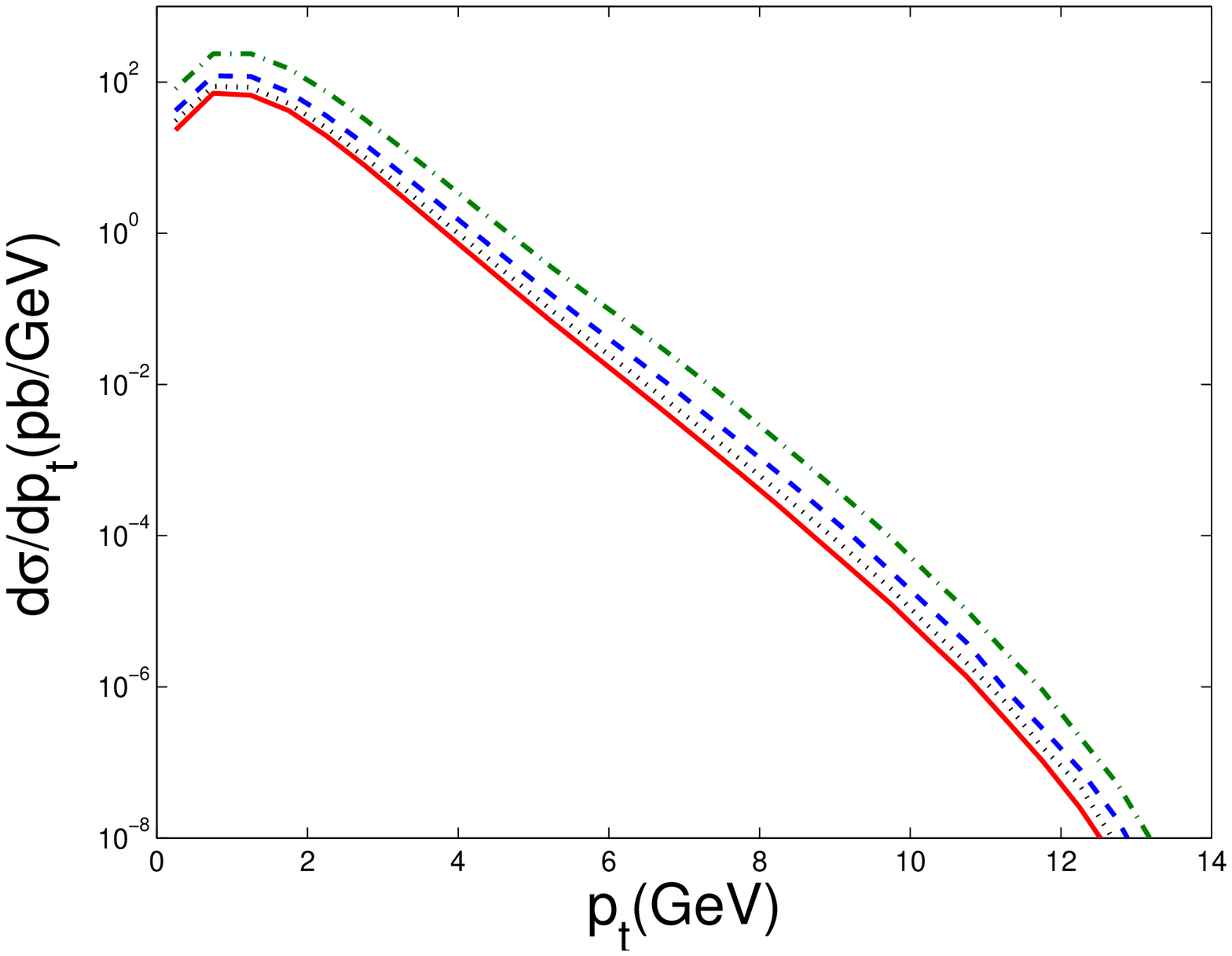}%
\hspace{0.5cm}
\includegraphics[width=0.46\textwidth]{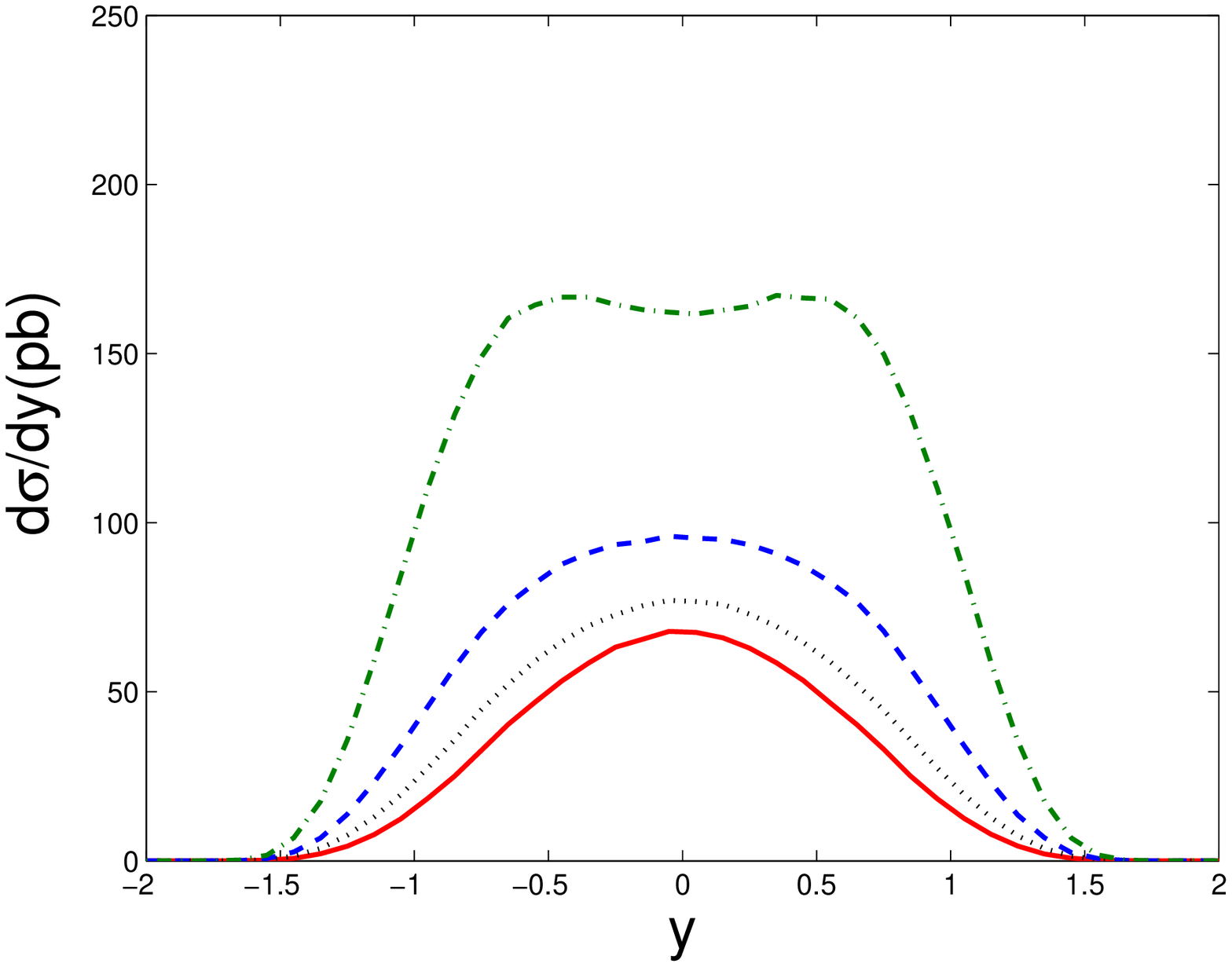}
\caption{The $p_t$-distributions (left) and $y$-distributions
(right) for the hadroproduction of $\Xi_{cc}$ at SELEX with
different values of $A_{in}$. The dotted, the dashed and the
dash-dot lines are for $A_{in}=0.1\%$, $0.3\%$ and $1\%$
respectively. The result with CTEQ6HQ, i.e., $A_{in}=0$ is shown by
a solid line (the lowest one).} \label{pty_varA}
\end{figure}

In Fig.\ref{pty_varA} we plot the $p_t$-distributions and the
$y$(rapidity)-distributions, where contributions from all the above
mentioned sub-processes are summed up. Our results show that the
$p_t$ distributions do not change their shapes significantly, but
the normalization changes. Also, as expected, the $p_t$ differential
cross-section becomes larger as $A_{in}$ increases from $0$ to
$1\%$. The shapes of $y$-distributions  and their normalization
change significantly by changing $A_{in}$  from $0$ to $1\%$. With
these results we may conclude that the intrinsic charm has
significant impact on the production of $\Xi_{cc}$ at SELEX.

\begin{table}
\begin{center}
\caption{The R values for SELEX with the cut $p_{t}>0.2$~GeV. }
\vskip 0.6cm
\begin{tabular}{|c||c||c|c|c|}
\hline - & {~~CTEQ6HQ($A_{in}=0$)~~}& {~~$A_{in}=0.1\%$~~}&
{~~$A_{in}=0.3\%$~~}& {~~$A_{in}=1\%$~~}\\
\hline\hline ~~$R$~~ & ~~$29.3$~~& ~~$36.6$~~
 &~~$51.3$~~ & ~~$103.$ ~~\\
\hline
\end{tabular}
\label{Rvalue}
\end{center}
\end{table}

Before the work of Ref.\cite{majp} and without taking the charm
PDF into account, the $\Xi_{cc}$ production at a hadron collider
is believed to be through the sub-process $gg\to (cc)_{\bf\bar
3}[^3 S_1]$.  Now with the extrinsic/intrinsic charm and the
$(cc)_{\bf 6}[^1 S_0]$ configuration contributions, there are
additionally several other sub-processes which are non-negligible. To
see how this alters the theoretical prediction based on the $gg\to
(cc)_{\bf\bar 3}[^3 S_1]$ subprocess, we introduce:
\begin{equation}\label{Rdef}
R=\frac{\sigma_{total}}{\sigma_{gg\to\Xi_{cc}
((cc)_{\bf\bar{3}}[^3S_1])}}\  ,
\end{equation}
where $\sigma_{total}$ stands for the cross section with
contributions from all sub-processes, including the intrinsic charm
initiated ones; ${\sigma_{gg\to\Xi_{cc}
((cc)_{\bf\bar{3}}[^3S_1])}}$ is the cross section only from the
sub-process $gg\to (cc)_{\bf\bar 3}[^3 S_1]$. The numerical results
of $R$ for SELEX is given in TABLE.\ref{Rvalue}. The results in
TABLE.\ref{Rvalue} show that the cross section at SELEX will be
enhanced by including all the dominant sub-processes contributions.
The enhancement can be even up to orders of magnitudes, e.g. the
prediction for the production cross-section can be enhanced by order
of about $10^2$ for the case of $A_{in}=1\%$. Hence it is possible
to reduce the discrepancy between theory and experiment from the order
of $10^3$ to $10^1$.

\begin{figure}
\centering
\includegraphics[width=0.50\textwidth]{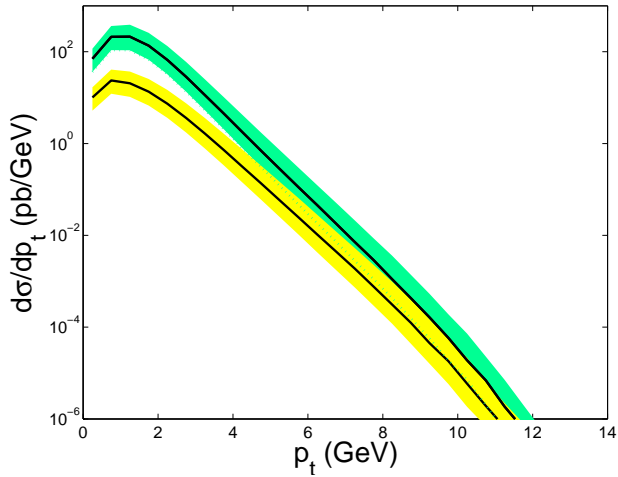}
\caption{The energy scale dependence of the $p_t$-distributions for
$g+c$ mechanism at SELEX. The upper band is for the case of
$(cc)_{\bf\bar 3}[^3S_1]$ and the lower band is for the case of
$(cc)_{\bf 6}[^1S_0]$, where the solid line in each band corresponds
to $\mu=M_t$, the upper edge of the band is for $\mu=M_t/2$ and the
lower edge is for $\mu=2M_t$. Here the intrinsic charm has been
taken into consideration, i.e. the PDFs of Eq.(\ref{apsg}) with
$A_{in}=1\%$ are used.} \label{selexscale}
\end{figure}

Finally, we give some results of  the uncertainty introduced by the
factorization/renormalization scale. For clarity, we take the
factorization scale $\mu_F$ and the renormalization scale $\mu_R$ to
be the same, $\mu_{F}=\mu_{R}=\mu$, and take three typical choices
for $\mu$ \cite{scale}, i.e. $\mu=M_{t}$ (the default one in our
calculations and $M_t\equiv \sqrt{M^2+p_{t}^2}$), $\mu=2M_{t}$ and
$\mu=M_{t}/2$. The energy scale dependence of the
$p_t$-distributions for $g+c$ mechanism are shown in
FIG.\ref{selexscale} for SELEX. Numerically, one may find that by
taking $\mu=M_t/2$ (or $\mu=2M_t$), the integrated cross-sections of
the $g+c\to\Xi_{cc}$ mechanism (either for the $(cc)_{\bf\bar
3}[^3S_1]$ configuration or the $(cc)_{\bf 6}[^1S_0]$ configuration)
will be increased (or decreased) by about $(2.0\sim4.0)$ times to
the case of $\mu=M_t$ within the allowable region of $p_t$.

\subsection{Hadronic Production of $\Xi_{cc}$ at  Tevatron and LHC}

\begin{figure}
\centering
\includegraphics[width=0.46\textwidth]{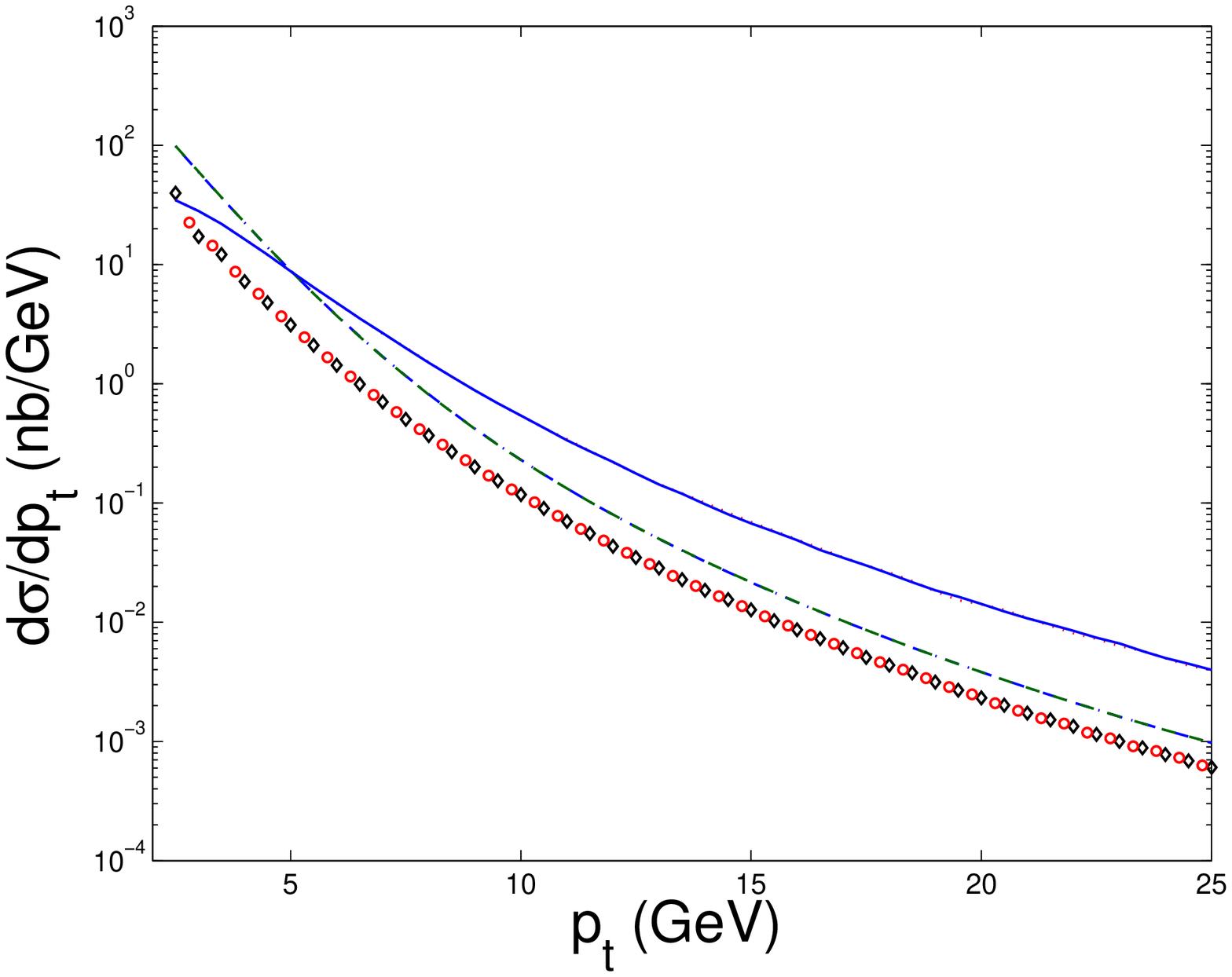}%
\hspace{0.5cm}
\includegraphics[width=0.46\textwidth]{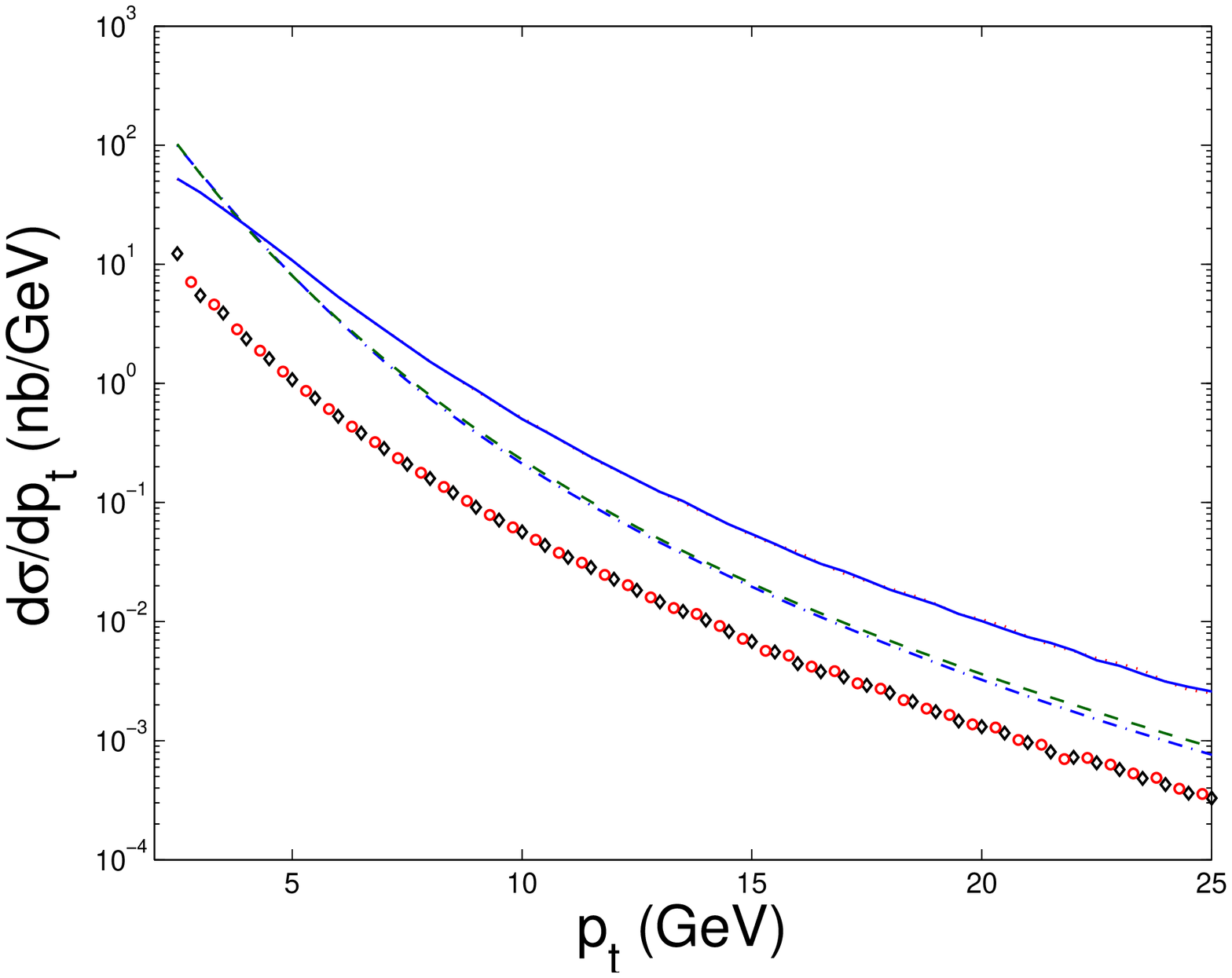}%
\caption{The $p_t$-distributions for the hadroproduction of
$\Xi_{cc}$ at LHC. The left figure is for CMS or ATLAS with the
rapidity cut $|y|\leq 1.5$ being adopted and the right one is for
LHCb with the pseudo-rapidity cut $1.8\leq|\eta|\leq 5.0$ being
adopted. The solid line, the dash-dot line and the circle line
correspond to that of the $g+g$, $g+c$ and $c+c$ mechanisms without
the intrinsic charm being considered (the PDFs in CTEQ6HQ
\cite{cqww} are used) respectively. The dotted line, the dashed line
and the diamond line correspond to that of the $g+g$, $g+c$ and
$c+c$ mechanisms with the intrinsic charm being considered (the PDFs
of Eq.(\ref{apsg}) with $A_{in}=1\%$ are used) respectively. The
differences with and without intrinsic charm are so small, that, of
them, only at LHCb for the $g+c$ mechanism the difference can be
seen from the right figure.} \label{lhcpt}
\end{figure}

\begin{figure}
\centering
\includegraphics[width=0.46\textwidth]{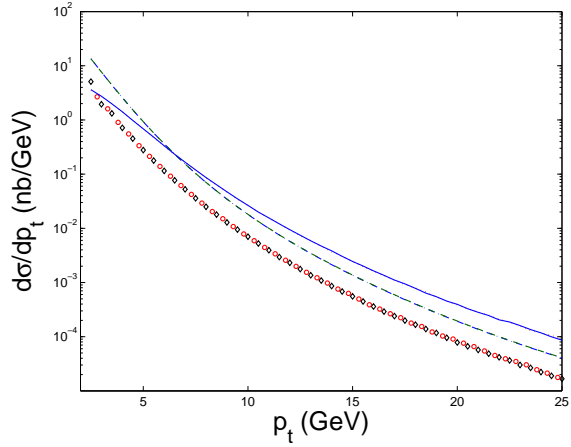}
\caption{The $p_t$-distributions for the hadroproduction of
$\Xi_{cc}$ at TEVATRON with the rapidity cut $|y|\leq 0.6$ being
adopted. The meaning for the lines in the figure is the same as
FIG.\ref{lhcpt}. The differences between the two cases with and
without intrinsic charm are too small to be seen.} \label{tevpt}
\end{figure}

\begin{figure}
\centering
\includegraphics[width=0.46\textwidth]{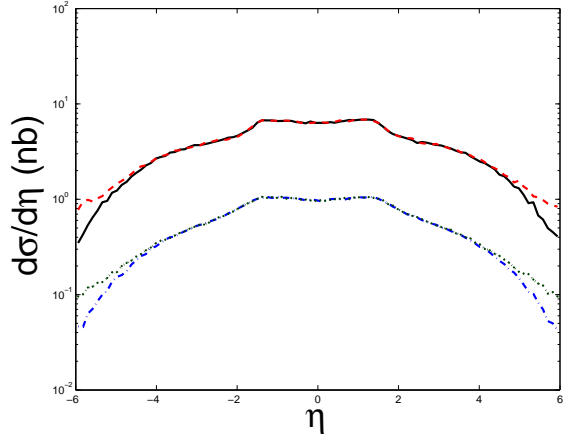}
\caption{The pseudo-rapidity $\eta$-distributions for the
hadroproduction of $\Xi_{cc}$ through the $g+c$ mechanism at LHC,
where the $p_t$ cut $p_t\geq 4$ GeV is adopted. The solid line, the
dash-dot line correspond to the case of $(cc)_{\bf\bar 3}[^3S_1]$
and the case of $(cc)_{\bf 6}[^1 S_0]$ without the intrinsic charm
being considered (the PDFs in CTEQ6HQ \cite{cqww} are used). The
dashed line and the dotted line correspond to the case of
$(cc)_{\bf\bar 3}[^3S_1]$ and the case of $(cc)_{\bf 6}[^1 S_0]$
with the intrinsic charm being considered (the PDFs of
Eq.(\ref{apsg}) with $A_{in}=1\%$ are used).} \label{lhcbgcy}
\end{figure}

We have evaluated the $\Xi_{cc}$ production at ATLAS or CMS and LHCb
at LHC (the center of mass energy $E_{cm}=14.\;{\rm TeV}$) and at
CDF or D0 at TEVATRON (the center of mass energy $E_{cm}=1.96\;{\rm
TeV}$), where the intrinsic charm contribution is taken into account
to see the possibility to observe the intrinsic charm effects at the
high energy colliders. Since at LHC, the detectors ATLAS and CMS are
similar, LHCb is quite different in rapidity cut; at TEVATRON the
detectors CDF and D0 are similar, thus we will present the results
bearing the similarity and difference in mind of the detectors. We
plot the $p_t$-distributions in FIGs.(\ref{lhcpt},\ref{tevpt}) with
or without the intrinsic of $A_{in}=1\%$, where the used  cuts in
rapidity are also given and the PDFs is taken from CTEQ6HQ
\cite{cqww} directly. From the figures one can see that for all of
the mechanisms $g+g$, $g+c$ and $c+c$ at the detectors of LHC and
TEVATRON, the differences in $p_t$-distributions for the two cases,
i.e., with or without the intrinsic charm are too tiny to be seen in
the most cases. The largest difference is found for the mechanism
$g+c$ at LHCb when $p_t$ is larger than $15$GeV (see the right
figure in FIG.(\ref{lhcpt})). In order to see this  more clearly, in
Fig.(\ref{lhcbgcy}), we plot the pseudo-rapidity distributions for
the g+c mechanism at LHC with a $p_{tcut}=4$ GeV. It can be found
that the difference for the two cases  is sizable only at large
pseudo-rapidity region.

\begin{table}
\begin{center}
\caption{Contributions from different sub processes to the cross
sections for the hadronic production of $\Xi_{cc}$ at Tevatron and
LHC. The cut $p_{t}\geq 4$~GeV is taken for all the hadronic
colliders. Additionally, the cut $|y|\leq 1.5$ is taken for LHC, the
cut $1.8\leq|\eta|\leq5.0$ is taken at LHCb, and the cut $|y|\leq
0.6$ is taken at Tevatron.} \vskip 0.6cm
\begin{tabular}{|c||c|c||c|c||c|c||}
\hline - & \multicolumn{2}{|c||}{~~Tevatron~~} &
\multicolumn{2}{|c||} {~~LHC~~}& \multicolumn{2}{|c||}
{~~LHCb~~}\\
\hline\hline - & ~~$(cc)_{\bf\bar 3}[^3S_1]$~~ & ~~$(cc)_{\bf
6}[^1S_0]$~~ & ~~$(cc)_{\bf\bar 3}[^3S_1]$~~ & ~~$(cc)_{\bf
6}[^1S_0]$~~ &~~$(cc)_{\bf\bar 3}[^3S_1]$~~ & ~~$(cc)_{\bf
6}[^1S_0]$~~ \\ \hline ~~$\sigma_{gg}\;({\rm nb})$~~ & ~~$1.61$~~ &
~~$0.399$~~ & ~~$22.3$~~ & ~~$5.44$~~ & ~~$25.7$~~ & ~~$6.47$~~ \\
\hline ~~$\sigma_{gc}\;({\rm nb})$~~ &
~~$2.31$~~ & ~~$0.361$~~ & ~~$22.1$~~ & ~~$3.42$~~ & ~~$20.6$~~ & ~~$3.00$~~ \\
\hline ~~$\sigma_{cc}\;({\rm nb})$~~ & ~~$0.755$~~ & ~~$0.0435$~~
& ~~$8.75$~~ & ~~$0.478$~~ & ~~$3.18$~~ & ~~$0.169$~~ \\
\hline\hline
\end{tabular}
\label{totcross}
\end{center}
\end{table}

In TABLE.\ref{totcross} we show the contributions from different
sub-processes to the cross sections. Our results are given with cuts
for different colliders and are derived by taking $A_{in}=1\%$. From
our results one can see that at Tevatron and LHC both
$gg\to\Xi_{cc}$ subprocess and $gc\to\Xi_{cc}$ subprocess are
dominant, unlike the situation at SELEX, where only the sub-process
$gc\to\Xi_{cc}$ is dominant. This implies that the intrinsic charm
has no significant impact on the $\Xi_{cc}$ production at the
Tevatron and LHC.

\begin{table}
\begin{center}
\caption{The value of $\varepsilon$ as defined in Eq.(\ref{kappa})
for the hadronic production of $\Xi_{cc}$ at Tevatron and LHC. In
the calculation, the cut $|y|\leq 1.5$ is taken for LHC, the cut
$1.8\leq|\eta|\leq 5.0$ is taken for LHCb, and the cut $|y|\leq 0.6$
is taken for Tevatron.} \vskip 0.6cm
\begin{tabular}{|c||c|c|c||c|c|c||c|c|c||}
\hline - & \multicolumn{3}{|c||}{Tevatron} & \multicolumn{3}{|c||}
{LHC}& \multicolumn{3}{|c||} {LHCb}\\ \hline $p_{tcut}$ & $10$~GeV &
$15$~GeV &$20$~GeV & $10$~GeV & $15$~GeV &$20$~GeV
& $10$~GeV & $15$~GeV &$20$~GeV \\
\hline\hline $\varepsilon_{gc}\,(p_{tcut})$ & $\sim1.1\%$ &
$\sim1.2\%$ & $\sim2.2\%$ & $\sim0.1\%$ & $\sim0.2\%$
& $\sim0.5\%$ & $\sim6.1\%$ & $\sim10.\%$ & $\sim15.\%$\\
\hline $\varepsilon_{cc}\,(p_{tcut})$ & $\sim1.5\%$ & $\sim2.2\%$ &
$\sim4.3\%$ & $\sim0.4\%$ & $\sim0.5\%$ & $\sim0.6\%$
& $\sim0.8\%$ & $\sim2.0\%$ & $\sim3.5\%$\\
\hline\hline
\end{tabular}
\label{ptcross}
\end{center}
\end{table}

To see how intrinsic charm leads to different contributions in different
$p_t$-regions, we introduce
\begin{equation}\label{kappa}
\varepsilon_i\,(p_{tcut})=\frac{\sigma_i(p_t\geq
p_{tcut})-\sigma^0_i(p_t\geq p_{tcut})}{\sigma^0_i(p_t\geq
p_{tcut})}\times 100\%,
\end{equation}
where $i=cc$ and $i=gc$ denote the contributions from the
sub-processes $cc\to \Xi_{cc}$ and $gc\to \Xi_{cc}$, respectively.
$\sigma_i^0$ denotes the cross section without the intrinsic charm,
while $\sigma$ denotes what with $A_{in}=1\%$. The results for the
ratio with different cuts are given in TABLE.\ref{ptcross}. From
TABLE.\ref{ptcross}, one can see that by introducing intrinsic
charm, the maximal change in the cross sections under various cuts
is only about $10\%$. This leads us to conclude that it may be
impossible to observe the effects of intrinsic charm at Tevatron and
LHC through the $\Xi_{cc}$ production.

\begin{figure}
\centering
\includegraphics[width=0.46\textwidth]{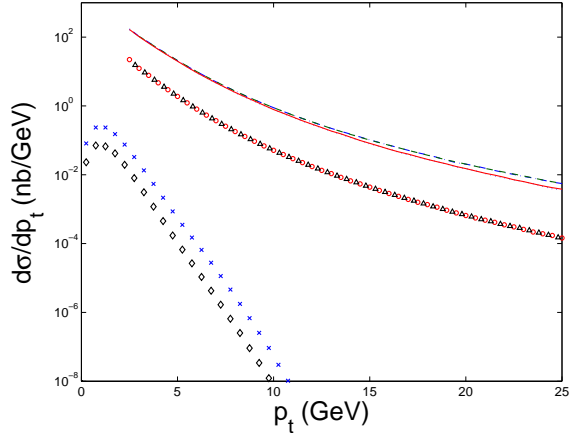}
\caption{The $p_t$-distributions for the hadroproduction of
$\Xi_{cc}$. The dotted line, the dash-dot line, the circle line and
the diamond line are those corresponding to LHCb, LHC, Tevatron and
SELEX respectively with $A_{in}=0$, respectively The solid, the
dashed line, the triangle line and the cross line are those
corresponding to LHCb, LHC, Tevatron and SELEX with $A_{in}=1\%$,
respectively. Only at SELEX, the difference between the cases with
and without intrinsic charm can be seen.} \label{sumpt}
\end{figure}

Finally we give in Fig.\ref{sumpt} the transverse distributions of
$\Xi_{cc}$ for different experiments. From Fig.\ref{sumpt} we see
that only at SELEX the effect of the intrinsic charm is significant
because the sub-process $gc\to\Xi_{cc}$ is dominant one.   At LHC
and Tevatron, the sub-process $gg\to\Xi_{cc}$ becomes more
important, and can even be dominant with the increasing $p_t$. This
is the main reason why our results for LHC and Tevatron do not show
any significant effect from the intrinsic charm.

\section{Discussions and Summary}

We have studied the $\Xi_{cc}$ hadronic productions in experiments
at SELEX, Tevatron and LHC. In our analysis we include contributions
from various sub-processes, especially contributions from the
intrinsic charm. For the intrinsic charm component inside a nucleon,
we adopt the BHPS model \cite{bhps}. Our results show that the
intrinsic charm can have significant impact on the $\Xi_{cc}$
production at SELEX. The cross section with certain reasonable
kinematic cuts can be four times larger than that without
considering the intrinsic charm. In contrast, the effect of the
intrinsic charm is small at LHC and Tevatron. The main reason is
that at those colliders with large energy, the contribution from the
sub-process of gluon-gluon fusion becomes bigger than that from the
sub-process of gluon-charm scattering (see TABLE.\ref{totcross}),
and the intrinsic charm only makes a small effect on the gluon PDF
(see Figs.\ref{methodA},\ref{methodAn}). Therefore, in comparison
with all of the environments considered here and if the intrinsic
charm component in a nucleon is really restricted in the region
$A_{in}=0.1\%$--$1\%$ as for the BHPS model, then only at SELEX the
$\Xi_{cc}$ production is suitable for `measuring' the intrinsic
charm component inside the incident hadrons. This is one of our main
results.

In the literature the hadronic production of $\Xi_{cc}$ is believed
to be dominated by the production of a $cc$ pair through gluon-gluon
fusion and the $cc$ pair is in a $[^3S_1]_{\bf\bar 3}$ state, and it
has been found that the predicted cross section based on this
sub-process is much smaller than the experimental observation of
SELEX, the discrepancy is roughly about $10^3$ \cite{comm}. The
transition probability of a $cc$ pair in $[^1S_0]_{\bf 6}$
configuration to $\Xi_{cc}$ is of the same order in $v$ as what the
$cc$ pair in $[^3S_1]_{\bf\bar 3}$ configuration \cite{majp}, so
that the extrinsic charm contribution can enhance the production
cross section quite a lot\cite{cqww}. Here we show that by taking
all the contributions from the intrinsic charm into account, the
prediction for the production cross-section can be enhanced by order
of about $10^2$ for the case of $A_{in}=1\%$, the discrepancy hence
is reduced. To obtain a final solution for the discrepancy a more
detailed study is needed.

\vspace{10mm}

\noindent {\bf\Large Acknowledgments:} This work was supported in
part by the Natural Science Foundation of China (NSFC). C.-F. Qiao
was supported also in part by the Scientific Research Fund of GUCAS
(NO. 055101BM03). X.-G. Wu thanks the support from the China
Postdoctoral Science Foundation.\\

\end{document}